\documentclass{PoS}

\title{Combined Dark Matter searches towards dwarf spheroidal galaxies with \emph{Fermi}-LAT, HAWC, HESS, MAGIC and VERITAS}

\ShortTitle{Combined DM Search FHHMV}

\author{\speaker{Louise Oakes}\\
        Humboldt University Berlin\\
        E-mail: \email{loakes@physik.hu-berlin.de}}
\author{Celine Armand\\
           LAPP, LAPTh\\
       E-mail: \email{celine.armand@lapth.cnrs.fr}}
\author{Eric Charles\\
        SLAC\\
        E-mail: \email{echarles@slac.stanford.edu}}
\author{Mattia di Mauro\\
        SLAC\\https://www.overleaf.com/5618749375mmyhwskgchnq
        E-mail: \email{dimauro.mattia@gmail.com}}
\author{Chiara Giuri\\
        DESY Zeuthen\\
        E-mail: \email{chiara.giuri@desy.de}}
\author{J. Patrick Harding\\
        Los Alamos National Laboratory \\
        E-mail: \email{jpharding@lanl.gov}}
\author{Daniel Kerszberg\\
        IFAE-BIST \\
        E-mail: \email{dkerszberg@ifae.es}}
\author{Tjark Miener\\
        IPARCOS, Universidad Complutense de Madrid \\
        E-mail: \email{tmiener@ucm.es}}
\author{Emmanuel Moulin\\
        CEA Saclay \\
        E-mail: \email{emmanuel.moulin@cea.fr}}
\author{Vincent Poireau\\
        LAPP \\
        E-mail: \email{poireau@lapp.in2p3.fr}}
\author{Elisa Pueschel\\
        DESY Zeuthen \\
        E-mail: \email{elisa.pueschel@desy.de}}
\author{Javier Rico\\
        IFAE-BIST \\
        E-mail: \email{jrico@ifae.es}}
\author{Lucia Rinchiuso\\
        CEA Saclay \\
        E-mail: \email{lucia.rinchiuso@cea.fr}}           
\author{Daniel Salazar-Gallegos\\
        Michigan State University \\
        E-mail: \email{salaza82@msu.edu}}            
\author{Kirsten Tollefson\\
        Michigan State University \\
        E-mail: \email{tollefson@pa.msu.edu}}            
\author{Benjamin Zitzer\\
        McGill University  \\
        E-mail: \email{bzitzer@physics.mcgill.ca}} 
\author{for the \emph{Fermi}-LAT Collaboration, the HAWC Collaboration \footnote{\texttt{http:\/\/www.hawc-observatory.org\/collaboration\/icrc2019.php}}, the H.E.S.S. Collaboration, the MAGIC Collaboration\footnote{\texttt{https://magic.mpp.mpg.de/}}~and the Veritas Collaboration.  \footnote{for all collaboration lists see PoS(ICRC2019)1177}}

\abstract{The search for Dark Matter (DM) has great potential to reveal physics beyond the Standard Model. As such, searches for evidence of DM particles are being carried out using a wide range of techniques, such as direct searches for DM particles, searches for DM produced with colliders, and indirect searches for the Standard Model annihilation products of DM.  Dwarf spheroidal galaxies (dSphs) are excellent targets for indirect Dark Matter searches due to their relatively high DM content and negligible expected astrophysical background.  A collaboration was formed to maximise the sensitivity of DM searches towards dSphs by combining for the first time dSph data from three imaging air Cherenkov telescope (IACT) arrays: HESS, MAGIC, and VERITAS; the \emph{Fermi}-LAT satellite, and the water Cherenkov detector HAWC. Due to the diverse nature of the instruments involved, each experiment will analyse their individual datasets from multiple targets and then the results will be combined at the likelihood level. For consistency of the likelihoods across the five experiments, a common approach is used to treat the astrophysical factor (J-Factor) for each target and an agreed set of annihilation channels are considered. We also agree on a common statistical approach and treatment of instrumental systematic uncertainties. The results are presented in terms of constraints on the velocity-weighted cross section for DM self-annihilation as a function of the DM particle mass.}

\FullConference{36th International Cosmic Ray Conference -ICRC2019-\\
		July 24th - August 1st, 2019\\
		Madison, WI, U.S.A.}

\begin{document}

\section{Introduction}
Searching for evidence of physics beyond the Standard Model (SM) has been a priority in science for several decades. A particularly promising field in which to find a hint of new physics is the search for Dark Matter (DM), to explain, for example, the significant deviation between the amount of observed matter in galaxies and the amount predicted as necessary to match their rotation parameters. Standard matter and energy is thought to make up only 20\% of matter in the Universe, with the rest consisting of DM. Searches for DM are carried out with a range of experimental techniques: \textit{direct detection} of interactions with DM particles through nuclear recoil; \textit{indirect detection} of the SM decay or annihilation products of DM particles; and searches for DM produced in \textit{accelerator experiments} such as at the CERN LHC. 

This talk focuses on the indirect detection techniques, using the observations of gamma rays from dwarf spheroidal galaxies (dSphs) with five different instruments. As gamma rays are neutral particles, they are not diverted by magnetic fields, so the incident direction leads back to the original source of the gamma rays. The choice of dSphs as targets was motivated by their high predicted DM density, very high mass to luminosity ratios and negligible astrophysical background making them an ideal laboratory for indirect DM searches~\cite{sarkar}.  Each of the experiments has published results of combinations of their own observed targets showing improved sensitivity over the individual results, leading naturally to this project to combine data from all five experiments. The large amount of observation data collected by the participating experiments on these sources is also a contributing factor.

Three of the experiments included are imaging atmospheric Cherenkov telescope (IACT) arrays: HESS, MAGIC, and VERITAS.  The others are HAWC, a water Cherenkov array, and \emph{Fermi}-LAT, a satellite telescope.  A collaboration was formed in 2018 to combine results from these five experiments for the first time, to optimise the sensitivity of DM searches by maximising the available statistics. MAGIC and \emph{Fermi}-LAT published together a joint paper~\cite{MAGICFermi} combining their dSph likelihoods to improve sensitivity to DM, which inspired this project to combine data from a wider range of experiments.  

From each of the five experiments, data from observations of several sources are included. Each experiment performed the analysis of their own data using agreed DM models, to avoid the need for sharing low-level experimental data and instrument response functions (IRFs) outside of the collaborations, and to allow for the appropriate handling of the different energy and angular resolutions and sensitive energy ranges of the various detector types. As no significant excess was observed for any source, upper limits were calculated and the likelihood functions for the velocity weighted cross section for the DM self-annihilation as a function of the DM particle mass for each source were combined.

The predicted flux of gamma rays from DM annihilation is given by Eqn.\ref{eqn:flux}\cite{bergstrom}:
\begin{equation}
\frac{d\Phi_{\gamma}}{dE_{\gamma}} = \frac{1}{4\pi}\frac{\langle \sigma_{ann} v\rangle} {2m^2_{DM}}\frac{dN_{\gamma}}{dE_{\gamma}} \times \int\limits_{d\Omega}d\Omega ' \int\limits_{los}\rho^2 dl(r,\theta ').
\label{eqn:flux}
\end{equation}
The observed photon flux is stated by the term $\frac{d\Phi_{\gamma}}{dE_{\gamma}}$ on the left-hand side. On the right-hand side the equation can be divided into two parts; the \textit{particle physics factor} dependent on  the velocity weighted cross section for the DM self-annihilation, $\langle \sigma_{ann} v\rangle$, the mass of the annihilating DM particles, $m_{DM}$, and the differential yield of gamma rays produced in an annihilation event , $dN_{\gamma}/dE_{\gamma}$; and the \textit{astrophysical factor}, here referred to as the \textit{J-factor}.  The J-factor is the integral over the line-of-sight of the dark matter density squared, $\rho^2$. 

Details of the individual experiments, datasets, analysis techniques and results are given in the following sections.

\section{Experiments and data}
\subsection{\emph{Fermi}-LAT}
The Fermi Large Area Telescope (LAT) is a satellite instrument operating since 2008. It is sensitive to gamma rays in the high energy range (HE) from 20 MeV to 300 GeV, covering the lowest energy region of the experiments in this study.  The LAT has a wide field of view covering about 20\% of the sky at a time and scans the whole sky approximately every 3 hours. The instrument detects HE gamma rays via pair conversion of the incoming photon to an electron and positron within the detector. Similarly to particle accelerator experiments, the direction of these particles is recorded in a silicon strip tracker and the energy deposited in a cesium iodide calorimeter~\cite{fermi2}.  An overview of the dSphs observed by \emph{Fermi}-LAT and a search for DM using the analysis of their combined data is given in~\cite{fermi2}.
\subsection{HAWC}
The High Altitude Water Cherenkov (HAWC) gamma-ray observatory is built on the Sierra Nevada volcano at an altitude of 4100 m, near Puebla, Mexico~\cite{hawc}. It is made up of 300 water Cherenkov detectors and is sensitive to gamma rays in the energy range 300 GeV to 100 TeV. The HAWC observatory detects showers of charged particles produced by TeV gamma rays as they interact with the Earth's atmosphere. These charged particles emit Cherenkov radiation as they travel through the water in the detectors. The radiation is detected using four photo-multiplier tubes (PMTs) per detector. The field of view of HAWC covers 15 \% of the sky at any time. The latest HAWC combined dSph search for DM is published in~\cite{hawc2}. HAWC performed an extended source analysis for the combination presented here instead of a point source analysis done in previous publications~\cite{hawc2}.

\subsection{Imaging Atmospheric Cherenkov Telescopes}
The following three experiments are IACTs two of which, MAGIC and VERITAS, are in the Northern, and the third, HESS is in the Southern Hemisphere. The Cherenkov detection technique is common to the three experiments. A shower of secondary particles is produced when a very high energy (VHE) gamma ray interacts with the Earth's atmosphere. These particles, travelling at close to the speed of light, stimulate the atmosphere to emit Cherenkov light in the direction of the original particle which triggered the shower. The intensity and direction of this Cherenkov light can be imaged by telescopes at ground level, which allows the reconstruction of the energy and incident direction of the source particle~\cite{hessweb}. The IACTs have fields of view in the range 3-5 degrees.  

\subsubsection{HESS}
The HESS observatory is an array of 5 IACTs situated at an altitude of 1800 m in the Khomas Highland of Namibia. The HESS phase I array consisted of four 12 m diameter telescopes. In 2012, phase II was implemented by the addition of a 28 m diameter telescope at the centre of the existing array. With the full phase II array HESS is can detect gamma rays with energies from 30 GeV to around 100 TeV. The data used in this analysis were collected with the phase I array between 2004 and 2012.  Previous HESS combined results of DM searches in dSph data can be found in~\cite{hessline} and ~\cite{hesscont}.
\subsubsection{MAGIC}
MAGIC consists of two IACTs, each with a diameter of 17 m, built on the Roque de Los Muchachos on the Canary Island of La Palma, at an altitude of 2200 m. The first telescope has been operational since 2004, the second was added to make a stereoscopic system in 2009. These telescopes are sensitive to VHE gamma rays in the range from tens of GeV up to several hundreds of TeV in normal observation mode. MAGIC have published their searches for DM towards dSphs in~\cite{magic2} and ~\cite{magic3}. 
\subsubsection{VERITAS}
VERITAS is an array of four IACTs of 12 m diameter, situated at the Fred Lawrence Whipple Observatory in Arizona, USA, at an altitude of 1268 m~\cite{veritas, veritas2} with the first telescope completed in 2004 and the last in 2007. The telescopes detect VHE gamma rays in the energy band from 100 GeV to 30 TeV. VERITAS results for DM searches towards dSphs, including a combination of their observed sources, are published in~\cite{veritas3}. 

\subsection{Datasets and individual analyses}
Table \ref{tab:stats} shows the details of the sources observed by each experiment. Data from these observations are analysed by the individual experiments, which then provide a likelihood function in an agreed format to be combined to produce the final upper limits.  The individual analyses are carried out using common J-factors and DM spectra, as well as the treatment of nuisance parameters, in order to produce comparable results between the experiments. The analysis details are given in the following section.
\begin {table}
{\footnotesize
\begin{center}
\begin{tabular}{ l|l } 
 \hline
 \hline
 Source name & Experiments \\
  \hline
 Bo\"otes I  & HAWC, VERITAS, \textit{Fermi}-LAT \\ 
 Canes Venatici I & \textit{Fermi}-LAT  \\ 
 Canes Venatici II & \textit{Fermi}-LAT, HAWC  \\ 
 Carina & HESS, \textit{Fermi}-LAT \\
 Coma Berenices & HAWC, HESS, \textit{Fermi}-LAT \\
 Draco &	HAWC, \textit{Fermi}-LAT \\
 Fornax &	H.E.S.S., \textit{Fermi}-LAT \\
 Hercules &	HAWC, \textit{Fermi}-LAT	\\
Leo I &	HAWC, \textit{Fermi}-LAT	 \\
Leo II	& HAWC, \textit{Fermi}-LAT	 \\
Leo IV & HAWC, \textit{Fermi}-LAT \\
Leo T & \textit{Fermi}-LAT \\
Leo V & \textit{Fermi}-LAT \\
Sculptor &	H.E.S.S., \textit{Fermi}-LAT \\
Segue I &	MAGIC, VERITAS, HAWC, \textit{Fermi}-LAT 	\\
Segue II & \textit{Fermi}-LAT \\
Sextans	& HAWC, \textit{Fermi}-LAT	 \\
Ursa Major I &	HAWC, \textit{Fermi}-LAT 	\\
Ursa Major II &	HAWC, MAGIC, \textit{Fermi}-LAT \\
Ursa Minor &	\textit{Fermi}-LAT	  \\
 \hline

\end{tabular}
\end{center}
}
\caption{Dwarf spheroidal galaxies included in the present combination of \textit{Fermi}-LAT, HAWC, HESS, MAGIC and VERITAS likelihoods. The right-hand column shows which experiments recorded the data for each source.  \label{tab:stats}}
\end{table}

\section{Analysis technique}
\subsection{Dark matter distribution and J-Factors}
DM profiles for the annihilation channels $\chi \chi$ to: $\tau^+ \tau^-$ and $ b \bar{b} $ are calculated according to~\cite{cirelli}. Further channels will be added later. J-factors are taken from~\cite{GS} and uncertainties on the J-factor are take into account in the likelihood as shown in the following section.
\subsection{Likelihood technique}
The joint likelihood function combining the measurements from each experiment and each dSph is:
\begin{equation}
    \mathcal{L}_{comb.}(\langle \sigma v\rangle; v |D) = \prod_{l=1}^{N_{dSph}}\mathcal{L}_{\gamma} (\langle \sigma v\rangle J_l;\mu_l | D_{\gamma ,l})\cdot \mathcal{L}_{J} (J_l| D_{J_l})
    \label{eqn:Lcomb}
\end{equation}

In~\ref{eqn:Lcomb} the likelihood depends on a set of nuisance parameters $v$ and is factorised into partial likelihood functions for each dSph, given by two terms. The first corresponds to the gamma-ray observations $\mathcal{L}_{\gamma}$, the second to the J-factor measurement $\mathcal{L}_{J}$. Within these terms, $\mu_l$ represents the nuisance parameters other than the total J-factor, $J_l$, $D_{\gamma ,l}$ is the gamma-ray data for the $l$-th dSph and $ D_{J_l}$ is the data constraining $J_l$. 

The likelihood term for the J-factor measurement, $\mathcal{L}_{J}$ is obtained from~\cite{GS}. It should be noted that most of the studied dSphs have been observed by more than one experiment, as can be seen in Table~\ref{tab:stats}. Therefore,  $\mathcal{L}_{\gamma}$ is defined for a specific dSph as the product over each set of observations for that source:

\begin{equation}
    \mathcal{L}_{\gamma} (\langle \sigma v\rangle J_l;\mu | D_{\gamma}) = \prod_{k=1}^{N_{meas}}\mathcal{L}_{\gamma ,k} (\langle \sigma v\rangle J_l;\mu_k | D_{\gamma ,k})
    \label{eqn:Lgamma}
\end{equation}
 where $\mu_k$ and $D_{\gamma, k}$ are the nuisance parameters and the gamma-ray data from the $k$-th measurement, respectively. The dSph index ($l$) has been removed in Eqn.~\ref{eqn:Lgamma} for clarity.
 
The values of $\mathcal{L}_{\gamma,k}$ are given as input for the combination by the individual experiments for each source they have observed. 

\section{Results}
\begin{figure}[ht]
\includegraphics[width=7cm]{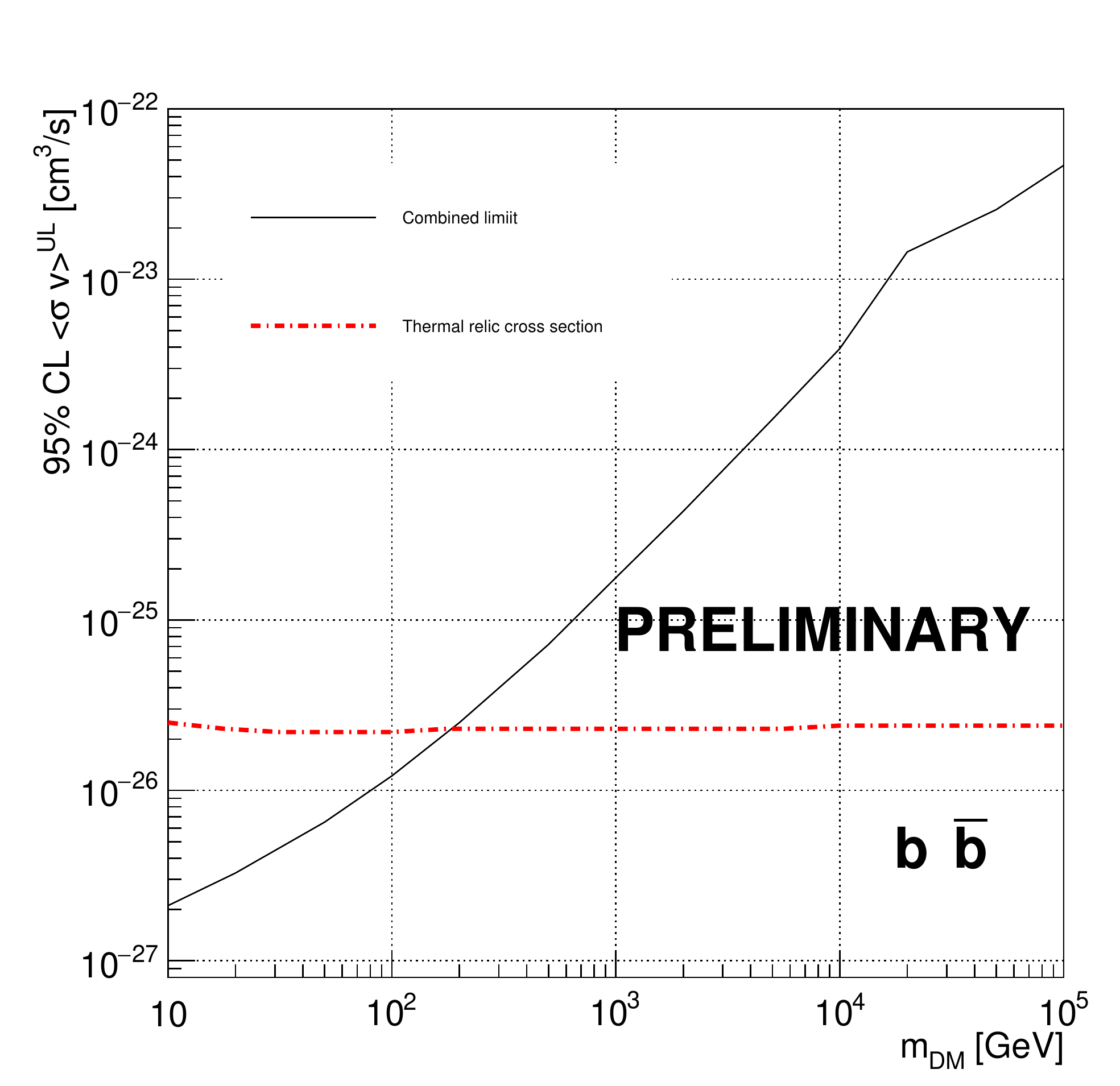}\includegraphics[width=7cm]{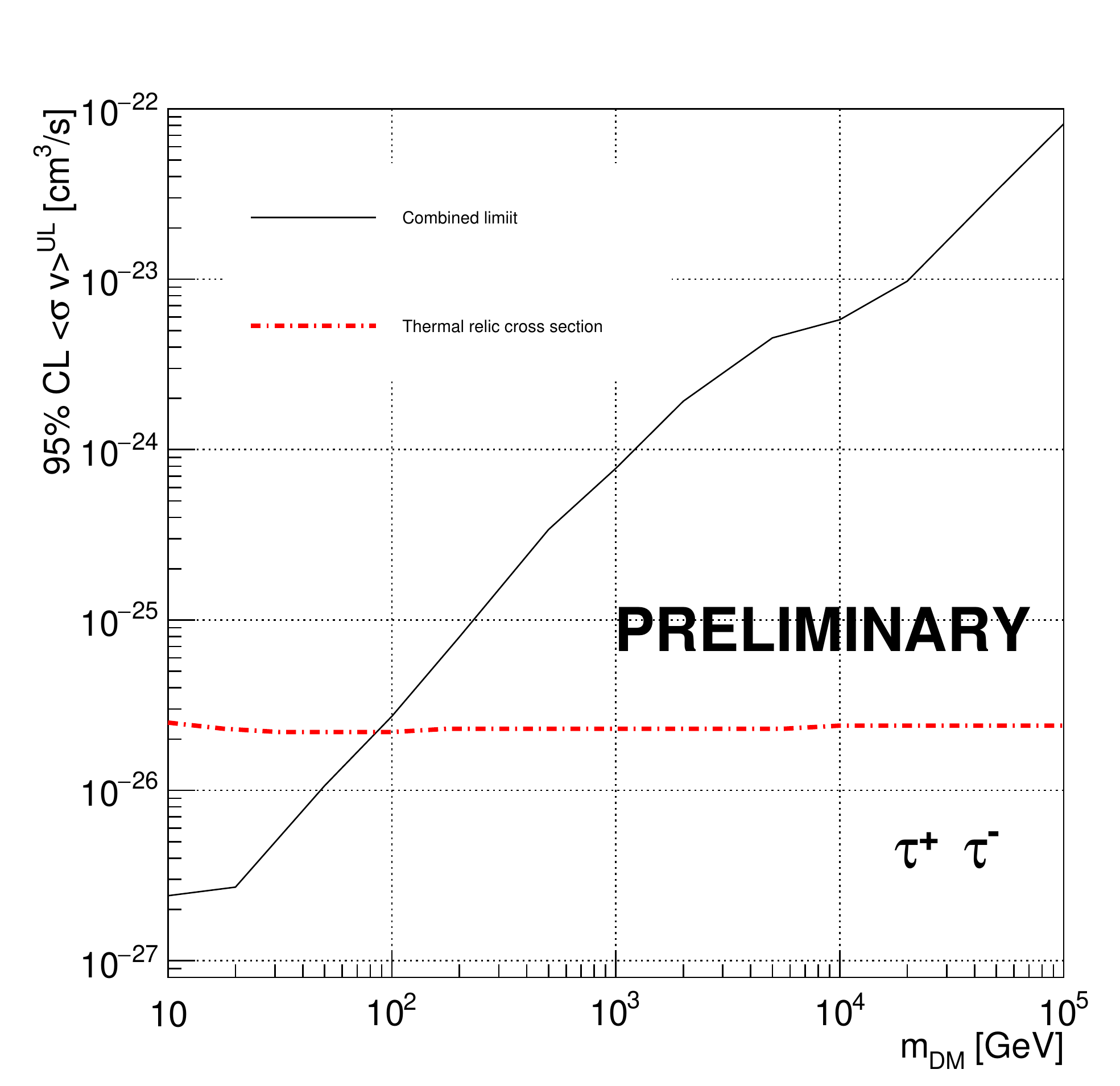}

\caption{95\% CL upper limits on the velocity weighted cross section,  $\langle \sigma v\rangle$, vs DM mass for DM particles annihilating into $ b\bar{b} $ (left) and $ \tau^{+}\tau^{-} $ (right). Results are from the combination of data from observations of 20 dSphs by \textit{Fermi}-LAT, HAWC, HESS, MAGIC and VERITAS. The thermal relic cross section is taken from~\cite{trc}.  \label{fig:UL}}
\end{figure}

Figure~\ref{fig:UL} shows a preliminary set of upper limits on the DM self-annihilation as a function of the DM particle mass for the combined data from 20 dSphs in the $ b \bar{b} $ and $\tau^+ \tau^-$ channels. Combining the limits results in an exclusion at the level of the thermal relic cross section of DM particles with masses below 100 GeV annihilating in the $ b \bar{b} $ channel and below 150 GeV in the $\tau^+ \tau^-$ channel.  
\section{Conclusions}
Preliminary results from the combination of data on 20 dSphs from \emph{Fermi}-LAT, HAWC, HESS, MAGIC and VERITAS have been shown. Data from individual instruments were provided in the form of likelihood functions in a common format by the individual instruments, then combined. No significant gamma-ray excess was observed after combination, so upper limits derived from the combined data have been presented.  The sensitivity of this combined analysis improves over the previously published results from the individual instruments, acting as a proof of principle for the full combination, which is in progress and will be published in the near future.  
\section{Acknowledgements}
\begin{itemize}
    \item HAWC full acknowledgements: http://www.hawc-observatory.org/collaboration/icrc2019.php  
    \item HESS full acknowledgements: https://www.mpi-hd.mpg.de/hfm/HESS/pages/publications/auxiliary/HESS-Acknowledgements-2019.html
    \item MAGIC full acknowledgements:  https://magic.mpp.mpg.de/acknowledgments\_ICRC2019\newline
    TM acknowledges support from the former {\em Spanish Ministry of Economy, Industry, and Competitiveness / European Regional Development Fund} grant FPA2015-73913-JIN.
    \item VERITAS:  This research is supported by grants from the U.S. Department of Energy Office of Science, the U.S. National Science Foundation and the Smithsonian Institution, and by NSERC in Canada. This research used resources provided by the Open Science Grid, which is supported by the National Science Foundation and the U.S. Department of Energy's Office of Science, and resources of the National Energy Research Scientific Computing Center (NERSC), a U.S. Department of Energy Office of Science User Facility operated under Contract No. DE-AC02-05CH11231. We acknowledge the excellent work of the technical support staff at the Fred Lawrence Whipple Observatory and at the collaborating institutions in the construction and operation of the instrument.
    \item \textit{Fermi}-LAT: The \textit{Fermi} LAT Collaboration acknowledges generous ongoing support from a number of agencies and institutes that have supported both the development and the operation of the LAT as well as scientific data analysis. These include the National Aeronautics and Space Administration and the Department of Energy in the United States, the Commissariat \`a l'Energie Atomique and the Centre National de la Recherche Scientifique / Institut National de Physique Nucl\'eaire et de Physique des Particules in France, the Agenzia Spaziale Italiana and the Istituto Nazionale di Fisica Nucleare in Italy, the Ministry of Education, Culture, Sports, Science and Technology (MEXT), High Energy Accelerator Research Organization (KEK) and Japan Aerospace Exploration Agency (JAXA) in Japan, and the K.~A.~Wallenberg Foundation, the Swedish Research Council and the Swedish National Space Board in Sweden. Additional support for science analysis during the operations phase is gratefully acknowledged from the Istituto Nazionale di Astrofisica in Italy and the Centre National d'\'Etudes Spatiales in France. This work performed in part under DOE Contract DE-AC02-76SF00515.
\end{itemize}

\end{document}